\documentclass[prb,amsmath,amssymb, 12pt]{revtex4}

\usepackage{graphicx}
\usepackage{bm}
\begin{document}

\title{Topologically stable states of the geometric quantum potential} 

\author{ Victor Atanasov}
\affiliation{Faculty of Physics, Sofia University, 5 blvd J. Bourchier, Sofia 1164, Bulgaria}

\author{Rossen Dandoloff}
\affiliation{Scientia Urbi Serviat, N.Kokanova str.1A, Sofia 1415, Bulgaria}


\begin{abstract}

We map the geometric quantum potential on the nonlinear sigma model and use homotopy to estimate the lower bound of the geometric quantum potential. 
We investigate a catenoid (wormhole section), a two dimensional bilayer geometry smoothly connected by a neck and a torus to show that in all these cases the geometric quantum potential creates topologically stable quantum states. 

\end{abstract}

\maketitle

A quantum particle constrained to abide a curved space or curved surface embedded in either curved or flat space, is no longer free. A curvature-induced quantum potential arises. Its origin can be traced to the operator ordering issue $\hat{r} \hat{p} \hat{r}^{-1} \hat{p} \neq \hat{p}^2$ successufully resolved by Podolsky with the introduction of the Laplace-Beltrami operator as the quantum mechanical kinetic energy\cite{podolsky}. Later DeWitt found that the Hamiltonian of the quantum system in curved space exhibits an extra term containing the Ricci scalar curvature\cite{dewitt}. The only ambiguity in his Feynman-inspired approach being the undetermined numerical factor to stand in front of $\hbar^2 R/m,$ in the sense that all choices lead to the  same classical theory. Again, the curvature-induced quantum interaction is compulsively resemblant to a kinetic term\cite{jj}.

The kinetic term origin of the curvature-induced quantum potential was confirmed in the thin layer confining potential approach championed by Jensen and Koppe\cite{JK} and daCosta\cite{daCosta}. The physical consequences are investigated in \cite{all}.

Let us note that the geometric quantum potential (GQP) has received its experimental verification in the context of photonic crystals\cite{photonic}, the electronic properties of one-dimensional metallic $\rm{C}_{60}$ polymer\cite{C60}. Therefore, the scope of the application of the geometric quantum potential is ubiquitous due to developments in nanotechnology\cite{AFS} which made it possible to grow quasi-two-dimensional surfaces of arbitrary shape where
quantum and curvature effects play a major role\cite{MT}. Examples
include single crystal $\rm NbSe_3$ M\"obius strips\cite{TTOIYH},
spherical $\rm CdSe-ZnS$ core-shell quantum dots\cite{SWFEB},
$\rm Si$ nanowire, nanoribbon transistors\cite{DNSPEG}, quantum
waveguides\cite{LCM}, nanotorus\cite{torus} and graphene. The list is not exhaustive, but deserves to end with the mention of carbon nanotubes and their versatile lengths, diameters, wall thicknesses and spatial thread-like configurations with which almost any space curve can be realised.

The geometric quantum potential (GQP) has the form\cite{daCosta}
\begin{equation}\label{daCosta}
V=-\frac{\hbar^2}{8m}(\kappa_1 - \kappa_2)^2= -\frac{\hbar^2}{2m} \left( M^2 -K \right).
\end{equation}
Here $\kappa_i$'s are the principal curvatures of the surface, $\hbar$ is the Planck's constant and $m$ is the effective mass. Here $M$ is the Mean curvature and $K$ is the Gauss curvaure of a two-dimensional surface embedded in three-dimensional space. The investigations until now were concentrated on different geometries and the corresponding GQP. So far the link between the GQP and the topology of the underlying geometry has not been discussed. 

Here, we map the GQP on the nonlinear sigma model which will allow us to connect to topology. Topology has been widely used in condensed matter to classify normalised vector fields, defects etc. \cite {RD} We use a widely unknown link between the normal $\bf n$ of a surface and its Gaussian and Mean curvatures. Indeed \cite{CEW}
\begin{eqnarray} \label{M_n}
 M &=& -\nabla\cdot{\bf n}, \\ \label{K_n}
 2K &=& {\bf n}\cdot \nabla ^2 \; {\bf n} +(\nabla \cdot {\bf n})^2.
 \end{eqnarray}
Using these expressions for $M$ and $K,$ in addition to
\begin{eqnarray} \label{M_n}
{\bf n}^2=1 &\Rightarrow& {\bf n}\cdot \nabla ^2 \; {\bf n} =-(\nabla {\bf n})^2
 \end{eqnarray}
in (\ref{daCosta}) we can transform  the GQP as follows:
\begin{eqnarray}\label{V_n}
V&=&-\frac{\hbar^2}{2m}(M^2 - K)=\\
&=&-\frac{\hbar^2}{2m} \left[  (\nabla\cdot{\bf n})^2 - 
\frac12 {\bf n}\cdot \nabla ^2 \; {\bf n} -\frac12 (\nabla \cdot {\bf n})^2
\right]=\\
&=&- \, \frac{\hbar^2}{4m} \left[(\nabla \, {\bf n})^2  + (\nabla \cdot {\bf n})^2\right].
\end{eqnarray}
We emphasise that

\begin{equation}\label{V&n}
V=  -\frac{\hbar^2}{2m}[(\nabla \, {\bf n})^2 + (\nabla \cdot {\bf n})^2]
\end{equation}
is the cornerstone realisation of the paper. Usually in the context of different spin models $(\nabla \, {\bf n})^2$ represents part of the energy density (in this case it represents the GQP plus the divergence squared term) which tends to 0 as ${\bf n} \rightarrow {\bf n}_0$, where ${\bf n}_0=\rm const$
denotes the constant vector filed at $\infty$. This behaviour of the vector field at infinity makes sure on one hand that the energy is finite and on the other hand allows a homotopical mapping onto $S^2$. We note here that ${\bf n}^2=1,$ that is the vector field has constant length. Let us note here that for a minimal surface ${\rm div} \, {\bf n} =0$\cite{CEW} and the expression for the GQP \label{V&n} matches perfectly the energy density for the nonlinear sigma model. In this case indeed the Gaussian curvature $K = (\nabla \, {\bf n})^2$.

\begin{figure}
\includegraphics[scale=0.45]{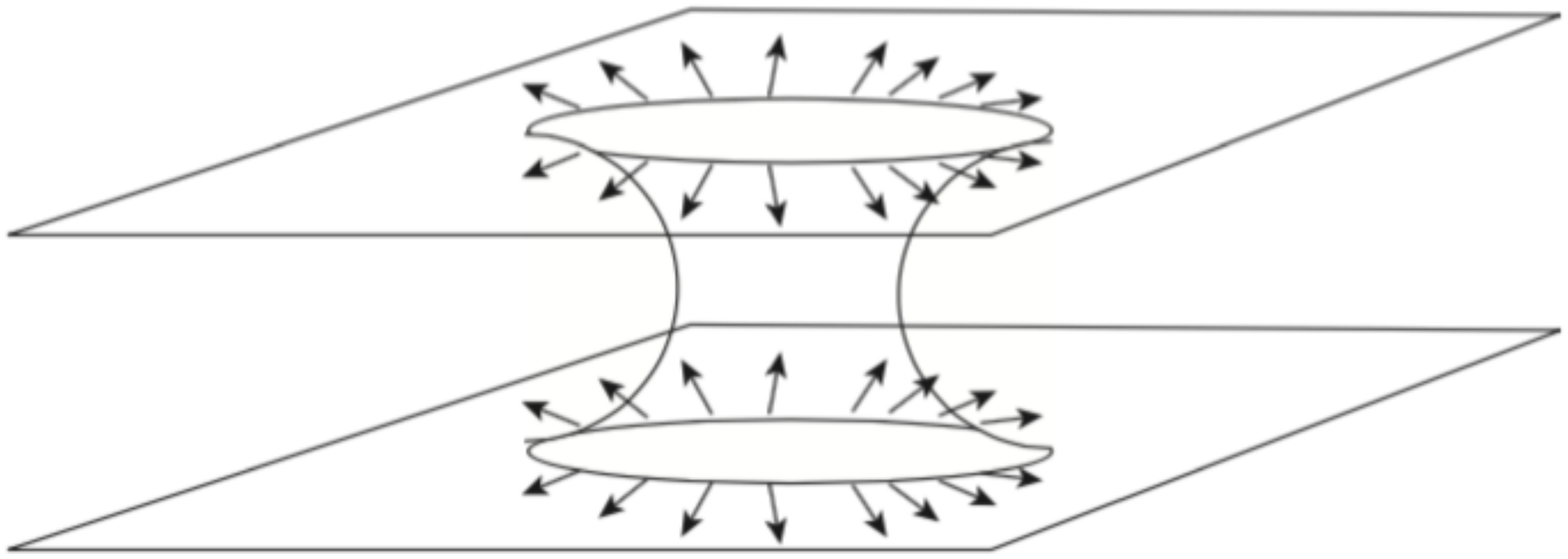}
\caption{\label{FIG1} Bilayer connected by a smooth neck.} 
\end{figure}

The first example we will consider here is the double layer connected by a smooth neck as shown on Figure \ref{FIG1}. A special feature of this surface is the fact that the first homotopy is different from zero: $\pi_1(Surface) \neq 0$ This automatically excludes the zero-th order of $\pi_2(Surface)$ where belongs the flat plane for which the GQP is vanishing. This surface can not continuously de deformed into the plane $R^2$.

The normals ${\bf n}$ to this surface at infinity on the upper layer are directed upwards and on the lower layer are directed downwards. When we map the normals from the upper layer and from the down layer onto the $S^2$ , we see that the vectors are sweeping the sphere $S^2$ once. In other words we can compactify the double layer with a neck  into a sphere $S^2$.

Now we will concentrate on the first class of the homotopy $\pi_2(S^2)$. In the case of a classical vector field (representing e.g. a classical spin field) with homogeneous boundary conditions
one can apply the Bogomolnyi inequality\cite{bog}:
\begin{equation}\label{bogomolnyi}
\int\int (\nabla{\bf n})^2dxdy \geq \int\int {\bf n}\cdot (\partial_x{\bf n}\wedge\partial_y{\bf n})dxdy = 4\pi
 \end{equation} 
 
The GQP (\ref{V_n}) is strictly negative and goes to zero at infinity. It is to be expected that there are localised states on this surface due to $V \leq 0$. Using the Bogomolnyi inequality we make an estimate concerning the minimal number of localised states on this geometry. The momenta in the potential well defined by the GQP vary between $0$ and $p$
 where $p=\sqrt{ -2mV}=\hbar\sqrt{(\nabla\cdot{\bf n})^2+(\nabla{\bf n})^2}$. The  density number of states $\cal N$ is given by:
 
 \begin{equation}
 {\cal N}=\frac{4\pi p^2}{(2\pi\hbar^2)^2}  = \frac{1}{\pi}((\nabla {\bf n})^2+(\nabla\cdot{\bf n})^2)\geq \frac{1}{\pi}(\nabla {\bf n})^2
 \end{equation}
 
Now, applying the Bogomolnyi inequality and integrating over the whole surface, we get for the total number of states the following inequality:
 \begin{equation}
  N=\int\int{\cal N}dS\geq\int\int\frac{1}{\pi}(\nabla {\bf n})^2dS \geq  \frac{1}{\pi} 4\pi = 4.
  \end{equation}
  So, the minimal number of states in this geometry created by the GQP and which are topologically stable, is 4.

\begin{figure}
\includegraphics[scale=0.45]{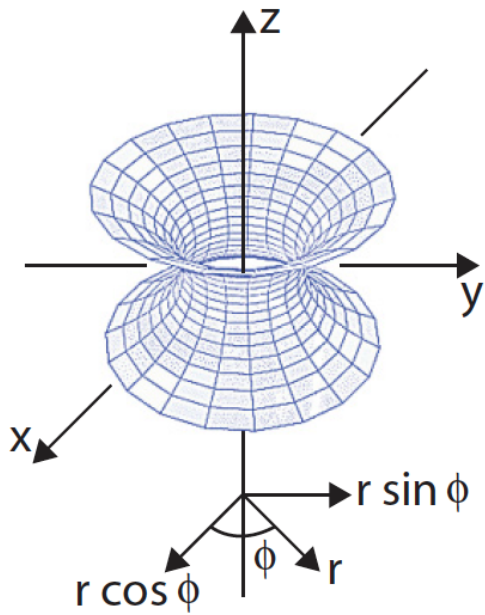}
\caption{\label{FIG2} Catenoid.} 
\end{figure}

 Our second example is the catenoid (shown on Figure \ref{FIG2}). Let us first note that at infinity the catenoid tends to the plane $\mathbb{R}^2$. This is best seen by considering the line element for the catenoid in polar coordinates, where $R$ is the radius of the throat of the catenoid \cite{rab}:
 \begin{equation}
  ds^2 = \frac{r^2}{r^2 - R^2}dr^2 + r^2d\phi^2
  \end{equation}
 Indeed when $r>>R$ the line element tends to the line element of the plane $\mathbb{R}^2$: $ds^2 = dr^2 + r^2 d\phi^2.$ In the upper half of the space at infinity the normals to the catenoid are pointing up and on the lower half at infinity they are pointing down, so that they can be mapped respectively to the north and south poles of the sphere $S^2$. Now we can map the normals to the catenoid to the sphere, covering it once. This is not surprising because the surface represented in Figure \ref{FIG1} can continuously be deformed (i.e. the vector field of the normals can also be deformed continuously) into the catenoid depicted on Figure \ref{FIG2}. In effect, the vector fields of the normals belong to the same homotopy class 1.
 The GQP for the catenoid (and for every minimal surface) is given by the energy density of the nonlinear sigma model:
  \begin{equation}
V=  -\frac{\hbar^2}{2m}[(\nabla \, {\bf n})^2]
\end{equation}  
Now we can directly apply the Bogomolnyi inequalities and get the number of topologically stable states created by the GQP which is 4.

Our last example will be the torus. The smooth degree 2 of mapping $T^2 \rightarrow S^2$ exists in contrast to the mapping $S^2 \rightarrow T^2$ which gives 0 (i.e. $\pi_2(T^2)=0)$. In the case of the mapping $T^2 \rightarrow S^2$ the normals to the torus when going from down to up passing from outside the torus and then coming down from inside the torus
sweep the sphere $S^2$ twice. As a result, now the minimum number of states created by the GQP will be 8.

Using the normal to a surface ${\bf n}$ as the main variable to express the Gaussian and the Mean curvatures we are able to map the GQP to the
nonlinear sigma model. This allowed for the use of techniques developed for the nonlinear sigma model and in particular the Bogomolnyi inequality in order to estimate the lower bound for the number of topologically protected bound states. We have explored the quantum potential for the catenoid, a two dimensional bilayer geometry smoothly connected by a neck and the torus. We believe the estimated lower number of topologically stable states for the GQP for these geometries can lead to a further nano-technological advances especially in devices exploring novel materials like graphene and graphene oxide.
 
V.A. acknowledges partial financial support by the National Science Fund of
Bulgaria under grant DN18/9-11.12.2017.

\end{document}